\documentclass[12pt]{article}
\usepackage{graphicx} 
\usepackage[top =2cm, bottom = 2cm, left= 2cm, right= 2cm]{geometry}
\usepackage{moreverb,url}
\usepackage{tikz}
\usepackage{multirow}
\usepackage[dvipsnames]{xcolor}
\usetikzlibrary{plotmarks}
\usepackage{booktabs}
\usepackage{authblk}
\usepackage{amsmath}
\usepackage{amsfonts}
\newcommand\BibTeX{{\rmfamily B\kern-.05em \textsc{i\kern-.025em b}\kern-.08em
T\kern-.1667em\lower.7ex\hbox{E}\kern-.125emX}}

\newcommand{\CI}{C\!I}

\author{Matilda Pitt \thanks{Email: mjp229@cam.ac.uk}}
\author{Robert Goudie}

\affil[1]{MRC Biostatistics Unit, University of Cambridge}
\begin{document}

\title{A Bayesian Prevalence Incidence Cure model for estimating survival using Electronic Health Records with incomplete baseline diagnoses}

\maketitle
\begin{abstract}
Retrospective cohorts can be extracted from Electronic Health Records (EHR) to study prevalence, time until disease or event occurrence and cure proportion in real world scenarios. However, EHR are collected for patient care rather than research, so typically have complexities, such as patients with missing baseline disease status. Prevalence-Incidence (PI) models, which use a two-component mixture model to account for this missing data, have been proposed. However, PI models are biased in settings in which some individuals will never experience the endpoint (they are ``cured''). To address this, we propose a Prevalence Incidence Cure (PIC) model, a 3 component mixture model that combines the PI model framework with a cure model. Our PIC model enables estimation of the prevalence, time-to-incidence, and the cure proportion, and allows for covariates to affect these. We adopt a Bayesian inference approach, and focus on the interpretability of the prior. We show in a simulation study that the PIC model has smaller bias than a PI model for the survival probability; and compare inference under vague, informative and misspecified priors. We illustrate our model using a dataset of 1964 patients undergoing treatment for Diabetic Macular Oedema, demonstrating improved fit under the PIC model. 
\end{abstract}
\section{Introduction}
When studying a disease, it is often of interest to estimate how many people have an outcome of interest at baseline (prevalence), how many people will go on to develop the outcome of interest and at what time (incidence) and how many people will never develop the outcome of interest (cure). These estimates can be used to form the foundation of comparison between different treatments, decide screening intervals, plan long term care and manage expectations for patients and clinicians \cite{ohno-machado_modeling_2001, spronk_calculating_2019} .

Furthermore, of increasing interest is using retrospective Electronic Health Records (EHR) to create datasets for analysis \cite{cowie_electronic_2017, casey_using_2016}. Compared to traditional clinical trial or cohort study datasets, EHR data can be cheaper and quicker to collect whilst covering more representative populations. However, there are additional challenges posed by EHR. For example, EHR often suffer from substantial amounts of missing data. Furthermore, we only have data when patients visit a health care provider, so data are interval-censored between appointments. Together, these complications make estimating prevalence and incidence cases non-trivial.

We focus on an application to people living with diabetes. Diabetes can cause damage to the retina in the eye, and a complication of this damage leads to Diabetic Macular Oedema (DMO) \cite{ciulla_diabetic_2003}. DMO leads to a decrease in eyesight, which we can measure as a Visual Acuity (VA) value on the Early Treatment of Diabetic Retinopathy Study charts \cite{shamir_comparison_2016}. Here, a `good' level of eyesight is defined as a VA value of 70 or above out of 100. There is evidence from randomised controlled trials (RCTs) that under a fixed interval-treatment regime eyesight can be improved with intravitreal injections of anti-vascular endothelial growth factor (VEGF) \cite{diabetic_retinopathy_clinical_research_network_phase_2007}. Strict adherence to this regime is rarely seen in real-life clinical settings, making the applicability of RCT evidence unclear. But, using EHR we can investigate prognosis under realistic, real-world treatment regimes. However, EHR data on DMO differ from RCT data in ways that need to be taken into consideration for a fair analysis. 

Two particular features of EHR data are present in our DMO dataset. Firstly, some patients may be `prevalent cases'. That is, they have a VA value $\geq 70$ at treatment initiation. This becomes an analytical challenge if some patients have no recorded baseline VA observation but have a VA value $\geq 70$ at their first subsequent appointment, meaning prevalent status cannot be ascertained. A second feature is that some patients may never attain a VA value $\geq 70$, suggesting they are `treatment non-responders' and that a cure model is required. 

Simple approaches to addressing these challenges do not provide reliable estimates. First, if patients with a missing prevalence status at baseline are assumed to be incident cases, the number of prevalent cases is underestimated. Furthermore, this approach gives biased estimates for the time until the incident disease outcome, because the missing prevalent cases will be incorrectly included in the incidence estimation. Second, if all patients are assumed to attain a VA value $\geq 70$, the survival function for incidence time will tend to a limit of zero, instead of tending to the probability of being a non-responder. Again, this results in biased survival estimates. Furthermore, this approach does not allow for analysis of whether covariates affect the probability of non-response, which would often be of scientific interest.

To address the challenge of unobserved prevalent cases, Prevalence-Incidence (PI) models have been proposed \cite{cheung_mixture_2017, hyun_flexible_2017}. Prevalence Incidence models use a two-component mixture model, with one component made up of those that are prevalent and the other component made up of those that are incident. The assignment of individuals to the two components is through a logistic model or probit model. The distribution of event times for incidence cases has been modelled via either a Weibull distribution, Cox model or an accelerated failure time model \cite{cheung_mixture_2017, hyun_flexible_2017}. Extensions for two-phase sampling, where some patients have additional covariates according to some sampling framework, have also been discussed \cite{hyun_sampleweighted_2020}. From the Bayesian approach, PI models have been further extended to include uncertainty in test results and allow for false negatives, motivated by a colonoscopy screening example \cite{klausch_bayesian_2025}. There, priors were placed on the sensitivity of test results and a Gibbs sampler derived. 

To address the challenge of treatment non-responders, cure models have been proposed. In some diseases a fraction of the population are `cured' (or non-susceptible or non-responders, depending on the event) of the event of interest, so will never experience it. For example, improvements in treatments mean that a proportion of patients with cancer can be considered cured after treatment, and will no longer die of cancer. However, which patients are cured is unknown at the completion of treatment \cite{othus_cure_2012}. Cure models, first introduced by Boag in 1949, assume that the `cured' part of the population will not experience the event of interest and estimate, with uncertainty, the proportion of `cured' patients \cite{boag_maximum_1949, othus_cure_2012, amico_cure_2018}. The most common approach is to model the population as a two-component mixture model, since the cure status of individuals is latent for all censored individuals. One component is made up of those who are susceptible (incident cases), who will experience the event, with event time modelled by a parametric or semi parametric distribution; and the second component is made up of those who are `cured', who will not experience the event \cite{de_la_cruz_bayesian_2022}. Standard cure models assume nobody in the cohort has the disease at baseline. 

Separate solutions to both challenges of DMO data are thus available. However, no existing approach can handle both challenges simultaneously.
While Prevalence Incidence models have been extended to allow for competing risks \cite{hyun_sampleweighted_2020}, the competing risk must be observable so it is not applicable in our setting, in which treatment non-response is unobservable.

We propose in this paper a Prevalence-Incidence-Cure (PIC) model that combines a Prevalence-Incidence model with a cure model to fully address the modelling challenges of DMO data that we have described. Our approach partitions the population into prevalent cases, incident cases and cured cases via a three-component mixture model. We describe how inference can be performed under the Bayesian paradigm, taking particular care when selecting priors to incorporate prior knowledge.

\section{Methods}
 
\subsection{PIC setup description}
We have a known number of individuals, $i=1, ..., n$, in our population sample at time $t=0$, with $t=0$ the baseline. This time could be, for example, the date a patient is first invited for a diagnostic test or the date of treatment initiation. We have a $q$-dim vector of covariates for each patient that is measured at baseline, $\boldsymbol{x}_{i} \in \mathbb{R}^q$. We can centre the covariates to help with interpretability as, for example, age at 0 does not always make sense to have as the base group. 

The event time of interest is $t_i$. This could be the time the disease first occurs, or the time a specific disease stage or threshold first occurs. We suppose we have some prevalent patients, who at baseline have already experienced the event of interest, meaning $t_i<0$. We also have some patients who are `cured' who will never experience the event of interest, meaning $t_i=\infty$. Depending on the application, `cured' could alternatively represent patients who are not susceptible to the disease, or who are non-responders to treatment, but we will hereafter refer to this group as `cured'.  

We assume that some patients have the diagnostic test at baseline $t = 0$ but not all, meaning we have baseline missingness, which occurs according to a missing at random process. We assume that we then observe whether the event of interest has occurred only at intermittent time points, when a diagnostic test is performed. For example, the data might arise from a population wide screening service or from patients receiving regular observation checks. We also assume that the time until the next diagnostic test is independent of the disease status of a patient. 

The true event time $t_i$ is therefore never directly observed due to interval censoring between diagnostic tests. We instead observe $l_i \in \mathbb{R}^+ \cup \{-\infty \}$ as the last time an individual is observed as disease free. When a patient tests positive at baseline, we have $l_i= -\infty$. We also observe $r_i \in \mathbb{R}^+ \cup \{\infty\}$ as the first time an individual is observed with the disease. When a patient is lost to follow up before experiencing the event or is `cured' then $r_i=\infty$. 

\begin{figure}
    \centering
    \includegraphics[width=1\linewidth]{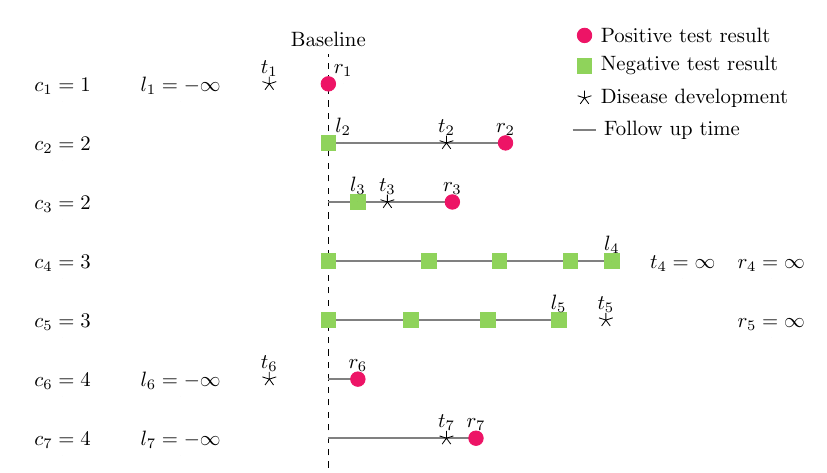}
    \caption{Illustrative examples of the disease and observation process for $7$ patients, including the true disease development time $t_i$ and the times of their diagnostic test. The times $l_i$ and $r_i$ of the last negative and first positive test respectively are also shown.}
   \label{diag: appointments}
\end{figure}

We introduce for each individual a variable $c_i \in \{1,2,3,4\}$ that partitions the data into four groups. The first group, with $c_i=1$, contains individuals with $r_i=0$, meaning they test positive at baseline. The second group, with $c_i=2$, contains individuals with $l_i \geq 0$ and $r_i< \infty$, meaning they are interval-censored incident cases. The third group, with $c_i=3$, contains individuals with $l_i \geq 0$ but $r_i = \infty$, meaning they are censored before they test positive. The final group, with $c_i=4$, contains individuals with $l_i = -\infty$ and $r_i > 0$, meaning they have a missing baseline test but test positive at their first appointment. An example of hypothetical patients with the times and outcome of their diagnostic tests, event development time and values for $r_i, l_i, t_i$ and $c_i$ category can be seen in Figure \ref{diag: appointments}.

\subsection{General prevalence-incidence-cure model}
We assume the true underlying process can be represented by a three-component mixture model with patients either prevalent, with probability $\pi_i= \mathbb{P}(\text{Patient $i$ is prevalent}| \boldsymbol{x}_i)$; incident cases, with probability $1-\pi_i-\delta_i$; or cured, with probability $\delta_i= \mathbb{P}(\text{Patient $i$ is cured}| \boldsymbol{x}_i)$. The survival function for incident cases is: 
\begin{equation}
    S(t_i|\boldsymbol{x}_i )= \mathbb{P}(T > t_i | \text{Incident}, \boldsymbol{x}_i )
\end{equation}
and is a proper survival function with $S(\infty)=0$. The corresponding probability density function is $f(t_i|\boldsymbol{x}_i )$. The complete likelihood where prevalence, cure status and true event time $t_i$ is known and including indicator functions $I()$ that take the value 1 if the argument is true and 0 otherwise, is given by:
\begin{equation}
\begin{split}
    L(\mathbf{t}, \mathbf{X}) = & \prod_{i=1}^n \pi_i^{I(i \text{ is prevalent})} \times\{(1-\pi_i-\delta_i) f(t_i|\boldsymbol{x}_i)\}^{I(i\text{ is incident}) }\times \delta_i^{I(i \text{ is cured})},
\end{split}
\end{equation}
with $\boldsymbol{X}=(\boldsymbol{x}_1, ..., \boldsymbol{x}_n)$ and  $\boldsymbol{t}=(t_1, ..., t_n)$.
The observed likelihood, however, accounts for unknown prevalence and cure status, and interval censored  event time between diagnostic tests. The observed likelihood is therefore given by:
\begin{equation}
\begin{split}
    L(\boldsymbol{l}, \boldsymbol{r},\boldsymbol{c},\boldsymbol{X}) = &
    \prod_{i=1}^n \pi_i^{I(c_i=1)} \times [(1-\pi_i-\delta_i)\{ S(l_i|\boldsymbol{x}_i)-S(r_i|\boldsymbol{x}_i)\}]^{I(c_i = 2)}\\
    & \times  \{\delta_i +(1-\pi_i-\delta_i) S(l_i|\boldsymbol{x}_i)\}^{I(c_i =3)}\\
    &\times [\pi_i + (1-\pi_i-\delta_i)\{1- S(r_i|\boldsymbol{x}_i)\}]^{I(c_i= 4)},
\end{split}
\label{eq: obslike}
\end{equation}
with $\boldsymbol{l}=(l_1, ..., l_n)$, $\boldsymbol{r}=(r_1, ..., r_n)$  and  $\boldsymbol{c}=(c_1, ..., c_n)$. The likelihood reflects only prevalence when $c_i=1$; and only incidence when $c_i=2$. When $c_i=3$, the likelihood represents the possibility that these individuals are either cured or are incident but censored. When $c_i=4$, the likelihood reflects that these individuals, had they been tested at baseline,  would either have tested positive (so are prevalent) or tested negative (so are incident).

\subsection{Multinomial Logistic-Weibull PIC model}
Our general PIC model does not specify the form of the prevalence probability $\pi_i$ or the cure probability $\delta_i$. We propose a multivariate logistic model for assignment of individuals to one of the three prevalent, incidence or cured components. The model parameters are $\pi_i, 1-\pi_i-\delta_i$ and $\delta_i$ defined in Eq. \ref{eq: pi} and Eq. \ref{eq: delta}. We define $\boldsymbol{x}_i^{\mathrm{mix}}=(1, \boldsymbol{x}_i')\in \mathbb{R}^{q^{\mathrm{mix}}}$, where $\boldsymbol{x}_i'$ is the subset of covariates relevant to the component-assignment model and $q^{\mathrm{mix}} \leq q+1$. Therefore, $\exp(\beta_{2\pi}),..., \exp(\beta_{q^{\mathrm{mix}}\pi})$ are the odds ratio of the covariate effects on prevalence and $\exp(\beta_{2\delta}),..., \exp(\beta_{q^{\mathrm{mix}}\delta})$ the odds ratio of the covariate effects on cure.
\begin{equation}
    \pi_i = \frac{\exp(\boldsymbol{\beta}_{\pi}\boldsymbol{x}_i^{\mathrm{mix}})}{\exp(\boldsymbol{\beta}_{\pi}\boldsymbol{x}_i^{\mathrm{mix}})+\exp(\boldsymbol{\beta}_{\delta}\boldsymbol{x}_i^{\mathrm{mix}})+1}
    \label{eq: pi}
\end{equation}
\begin{equation}
    \delta_i = \frac{\exp(\boldsymbol{\beta}_{\pi}\boldsymbol{x}_i^{\mathrm{mix}})}{\exp(\boldsymbol{\beta}_{\pi}\boldsymbol{x}_i^{\mathrm{mix}})+\exp(\boldsymbol{\beta}_{\delta}\boldsymbol{x}_i^{\mathrm{mix}})+1}
    \label{eq: delta}
\end{equation} 
We use a Weibull distribution as the parametric time-to-event distribution for incidence cases for illustrative purposes and due to its interpretable parametrisations and support from the theory of carcinogenesis \cite{armitage_age_2004}. The same covariates do not need to be in both the logistic model and the Weibull survival model, so let $\boldsymbol{x}_i^{\mathrm{inc}} \in \mathbb{R}^{q^{\mathrm{inc}}}$, with $q^{\mathrm{inc}} \leq q$, denote the covariates in the Weibull survival model. The Weibull model can be parametrised to give either a proportional hazards model or an accelerated failure time model. We use a proportional hazards parametrisation so the covariate effects can be described in terms of hazard ratios for incident disease, using parameter $\boldsymbol{\gamma} \in \mathbb{R}^{q^{\mathrm{inc}}}$. 
For proportional hazards with a Weibull baseline, the hazard function $h_i(t)$, including the shape parameter $\alpha \in \mathbb{R}$, and the scale parameter $\lambda\in \mathbb{R}$, is
\begin{equation}
    h(t_i| \boldsymbol{x}^{\mathrm{inc}}_i)=\frac{\alpha}{\lambda^{\alpha}}t_i^{\alpha-1}\exp(\boldsymbol{\gamma}\boldsymbol{x}^{\mathrm{inc}}_i),
\end{equation}
with corresponding survival function
\begin{equation}
\begin{split}
    S(t_i|\boldsymbol{x}^{\mathrm{inc}}_i) &= \exp\left[-\left\{\frac{t_i}{\lambda}\exp\left(\frac{\boldsymbol{\gamma}\boldsymbol{x}_i^{\mathrm{inc}}}{\alpha}\right) \right\}^{\alpha}  \right].
    \end{split}
    \label{eq: s}
\end{equation}

\subsection{Priors}
Here, we suggest informative priors for our multinomial logistic-Weibull model because in many medical scenarios we have knowledge of the disease event from previous research or at least knowledge of impossible parameter values that we want to include to reduce parameter uncertainty. For example, we may have knowledge from previous clinical trials that we expect to translate across to our EHR data from real-life clinical settings, as in the DMO case.

\subsubsection{Mixture model}
For the component-assignment model, we place priors on the ratio of the group allocation probabilities. More specifically we place priors on the ratio $r_{\pi}$ of the probability of being cured to being an incident case and the ratio $r_{\delta}$ of the probability of being a prevalent case to being an incident case. Elicitation is based on the ratios because we assume experts can convey their knowledge more easily through these comparisons, as opposed to directly on the $\beta_{1\pi}$ or $\beta_{1\delta}$ values, which refer to the log of the odds ratio. This closely matches the suggestion by Elfadaly and Garthwaite who developed software to allow for expert elicitation for multivariate logistic regression models \cite{elfadaly_quantifying_2020}. For models with covariates, the elicitation is made under the base group where all covariates are taken as 0 with $\boldsymbol{x}_0= \mathbf{0}$,  $\pi_0= \mathbb{P}(\text{Patient is prevalent}| \boldsymbol{x}_0)$ and $\delta_0= \mathbb{P}(\text{Patient is cured}| \boldsymbol{x}_0)$.
\begin{equation}
    r_{\pi}=\frac{\mathbb{P}(\text{Prevalent}| \boldsymbol{x}_0)}{\mathbb{P}(\text{Incident}| \boldsymbol{x}_0)}=\frac{\pi_0}{1-\pi_0-\delta_0}
    \label{eq: ratio pi}
\end{equation}
\begin{equation}
    r_{\delta}=\frac{\mathbb{P}(\text{Cured}|\boldsymbol{x}_0)}{\mathbb{P}(\text{Incident}| \boldsymbol{x}_0)}=\frac{\delta_0}{1-\pi_0-\delta_0}
    \label{eq: ratio delta}
\end{equation}
We use log-normal priors for these ratios
\begin{equation}
     r_{\pi} \sim \text{LogNormal}(\mu_{1\pi}, \sigma_{1\pi}^2)
\end{equation}
\begin{equation}
     r_{\delta} \sim \text{LogNormal}(\mu_{1\delta}, \sigma_{1\delta}^2)
\end{equation}

We specify the means $\mu_{\pi}, \mu_{\delta}$ and the scales $\sigma_{\pi}, \sigma_{\delta}$ by eliciting quantiles of the ratios, $r_{\pi}$ and $r_{\delta}$ respectively. The values of the 95\% quantiles for the ratios, $\CI_{97.5}$ and $\CI_{2.5}$ such that $\mathbb{P}(r<\CI_{97.5})=0.975$ and $\mathbb{P}(r<\CI_{2.5})=0.025$, are elicited from the expert or previous research and the following equations used to find the lognormal parameters, $\mu$ and $\sigma$:
\begin{equation}
    \mu=\frac{\log(\CI_{2.5})\times \Phi^{-1}(0.975)- \log(\CI_{97.5})\times\Phi^{-1}(0.025)}{2\times\Phi^{-1}(0.975)}
    \label{eq: mu elicit}
\end{equation}
\begin{equation}
    \sigma=\frac{\log(\CI_{97.5})- \log(\CI_{2.5})}{2\times\Phi^{-1}(0.975)}
    \label{eq: sigma elicit}
\end{equation}
with $\Phi()$ denoting the standard normal CDF. As $\beta_{1\pi}=\log(r_\pi)$ and $\beta_{1\delta}=\log(r_\delta)$ it follows that their priors are normally distributed:
\begin{equation}
    \beta_{1\pi}\sim \text{Normal}(\mu_{1\pi}, \sigma_{1\pi}^2)
\end{equation}
\begin{equation}
    \beta_{1\delta}\sim \text{Normal}(\mu_{1\delta}, \sigma_{1\delta}^2)
\end{equation}
For priors on the regression coefficients $\beta_{2\pi},...,\beta_{q^{\mathrm{mix}}\pi }$, prior elicitation is made on the $95\%$ quantiles for the odds ratio of the individual covariate effects on prevalence, $\exp(\beta_{2\pi}),...,\exp(\beta_{q^{\mathrm{mix}}\pi })$, using Eq. \ref{eq: mu elicit} and \ref{eq: sigma elicit}.
\begin{equation}
    \exp(\beta_{j\pi})\sim \text{LogNormal}(\mu_{j\pi}, \sigma_{j\pi}^2), \text{ }j= 2,..., q^{\mathrm{mix}}
\end{equation}

Similarly, for priors on the regression coefficients $\beta_{2\delta},...,\beta_{q^{\mathrm{mix}}\delta}$, prior elicitation is made on the $95\%$ quantiles for the odds ratio of the individual covariate effects on cure, $\exp(\beta_{2\delta}),...,\exp(\beta_{q^{\mathrm{mix}}\delta})$,  using Eq. \ref{eq: mu elicit} and \ref{eq: sigma elicit}.
\begin{equation}
    \exp(\beta_{j\delta})\sim \text{LogNormal}(\mu_{j\delta}, \sigma_{j\delta}^2), \text{ }j=2,..., q^{\mathrm{mix}}
\end{equation}
It follows a log transformation can be performed for priors directly to give:
\begin{equation}
    \beta_{j\pi} \sim \text{Normal}(\mu_{j\pi}, \sigma_{j\pi}^2), \text{ }j=2,..., q^{\mathrm{mix}},
\end{equation}
\begin{equation}
    \beta_{j\delta} \sim \text{Normal}(\mu_{j\delta}, \sigma_{j\delta}^2), \text{ }j=2,..., q^{\mathrm{mix}}.
\end{equation}

\subsubsection{Incident event time model}
For the Weibull distribution, the shape parameter $\alpha$ represents the incidence rate over time. We assume that prior studies or a clinician can provide prior insight into whether the incidence rate increases, $\alpha > 1$, or decreases over time, $\alpha < 1$. 
Since $\alpha>0$ we use a log-normal prior with the  prior elicitation made on the 95\% confidence intervals for $\alpha$ and using Eq \ref{eq: mu elicit} and \ref{eq: sigma elicit}. 
\begin{equation}
    \alpha \sim \text{LogNormal} (\mu_{\alpha}, \sigma_{\alpha}^2)
\end{equation}
The scale parameter, $\lambda$, is less intuitive than $\alpha$ so we instead make elicitation on the median time to event  $\widetilde{m} = \lambda \ln(2)^{\frac{1}{\alpha}}$. As median lifetime must also be greater than $0$, a log-normal distribution is used with the 95\% quantiles matching the prior elicitation. 
\begin{equation}
    \widetilde{m} \sim \text{LogNormal} (\mu_{\lambda}, \sigma_{\lambda}^2)
\end{equation}
The transformation $\lambda = \widetilde{m}\{ln(2)^{\frac{1}{\alpha}}\}^{-1}$ is then performed to obtain the prior for $\lambda$.
When including a covariates, the prior for median survival time should instead be specified for the base group, where $\boldsymbol{x}^{\mathrm{inc}}=\mathbf{0}$. A log-normal distribution for the hazard ratio of the covariate effect, $\exp(\boldsymbol{\gamma})=(\exp(\gamma_1),..., \exp(\gamma_{q^{\mathrm{inc}}}))$,  can be elicited for each covariate from previous literature or by consulting a clinician,
\begin{equation}
    \exp(\gamma_j) \sim \text{LogNormal} (\mu_{j\alpha}, \sigma_{j\alpha}),\text{ }j= 1,..., q^{\mathrm{inc}}.
\end{equation}
It follows that
\begin{equation}
    \gamma_{j} \sim  \text{Normal} (\mu_{j\alpha}, \sigma_{j\alpha}) , \text{ }j= 1,..., q^{\mathrm{inc}}.
\end{equation}
\subsubsection{Complete prior specification}
In summary, in the case with $q^{\mathrm{mix}}$ and $q^{\mathrm{inc}}$ covariates in the component-assignment and the time-to-event models, respectively, the specification is as follows.
\begin{equation}
    \begin{split}
        \alpha & \sim \text{LogNormal}(\mu_{\alpha}, \sigma_{\alpha}^2)\\
        \widetilde{m} & \sim \text{LogNormal}(\mu_{\lambda}, \sigma_{\lambda}^2)\\
        \beta_{j\pi}& \sim  \text{Normal}(\mu_{\beta_{j\pi}},  \sigma_{\beta_{j\pi}}^2), \text{ }j=1,..., q^{\mathrm{mix}} \\
        \beta_{j\delta}& \sim  \text{Normal}(\mu_{\beta_{j\delta}}, \sigma_{\beta_{j\delta}}^2), \text{ }j= 1,..., q^{\mathrm{mix}}\\
        \gamma_{j\delta}& \sim  \text{Normal}(\mu_{\gamma_{j\delta}}, \sigma_{\gamma_{j\delta}}^2), \text{ }j= 1,..., q^{\mathrm{inc}}\\
    \end{split}
\end{equation}

\subsection{Model estimation}
We implement our model in JAGS, which draws samples from the posterior distribution for the parameters via Markov chain Monte Carlo \cite{plummer_jags_2017, alvares_bayesian_2021}. To specify the observed likelihood, as seen in Eq. \ref{eq: obslike}, we use the ``zeros trick" approach  \cite{lunn_bugs_2012}. Full reproducible code is on GitHub at https://github.com/Tilly-Pitt/PICTutorialCode. Convergence was assessed using the Gelman-Rubin statistic, with a value $<1.1$ considered converged \cite{gelman_inference_1992}. The posterior medians are used as point estimates of the parameters.

\section{Simulation study}
There are two main aims of the simulation study. Our first aim is to assess the bias and coverage of parameter estimates when using our proposed model across a range of realistic settings, including with informative, vague and misspecified priors. Our second aim is to compare to an existing PI model according to estimates of prevalence, cure and survival probabilities. 

We design our simulation study to reflect characteristics of cervical cancer screening, for which PI models have previously been used \cite{cheung_mixture_2017, hyun_sampleweighted_2020}.
In this setting, as an initial screening test, patients are offered a human papillomavirus infection (HPV) test. If the HPV test is positive, the individual is considered high risk for cervical cancer. These high risk individuals form the cohort for our simulation study, and are offered either an immediate baseline diagnostic test (colposcopy) or regular retests to detect cancerous cells. Since not all high risk individuals have a diagnostic test at baseline, prevalence is unknown. Furthermore, if the HPV virus is cleared before precancer cells develop then the patient will no longer go on to develop cancer from that HPV infection and can be considered cured \cite{hyun_sampleweighted_2020}. For our simulation study, we assume the time of HPV infection clearance is unknown, precluding use of a competing risk model.

\subsection{Data generating process}
We chose the parameters of our simulation study to reflect the cervical cancer screening application as closely as possible \cite{cheung_mixture_2017, hyun_sampleweighted_2020}. In the original paper, those aged 24-44 were analysed separately from those over 45 \cite{hyun_sampleweighted_2020}. We used the estimates from those aged 24-44 to generate the data in our simulation study.

We simulated whether an individual has a baseline test by drawing samples from a Bernoulli random variable, with probability 0.88 of a baseline test. To simulate interval censoring, the time until subsequent diagnostic tests was sampled from a gamma distribution, with shape parameter 2 and scale parameter 1. The number of diagnostics tests per person was the minimum until either: their incidence event time had occurred and been observed; time 20 had been reached; or 10 diagnostic test times had been simulated.

We sampled prevalence, cure, incidence status from a multinomial distribution with parameters as seen in Eq. \ref{eq: pi} and Eq. \ref{eq: delta}. Cancer risk varies by strain of HPV, so we include a single binary covariate $x_i$ indicating whether the individual has HPV strain 16 or 18, which we assume is measured at baseline. In our simulations individuals are strain 16 with probability 0.76. In strain 16 we set the probability of prevalence $\pi=0.169$ and the probability of cure $\delta =0.755$, meaning $\beta_{1\pi} =0.799$ and $\beta_{1\delta} =2.296$. In strain 18 we set the probability of prevalence $\pi=0.081$ and the probability of cure $\delta = 0.756$, meaning $\beta_{2\pi} =-1.498$ and $\beta_{2\delta} =-0.762$. We set $\gamma=-0.5$ as strain 16 is known to cause more cancer than strain 18 \cite{munoz_epidemiologic_2003}.

We sampled event times for incidence cases from a Weibull distribution as seen in Eq. \ref{eq: s}. We set $\alpha=2$, $\lambda=2.2$ for the Weibull event times to give a median event time of 22 months. 

We considered three sample sizes:  $n=40$ reflecting an uncommon disease event with few patients, such as cervical cancer in pregnant women \cite{he_effect_2022}; $n=400$ reflecting a larger EHR dataset but within a subset of the population, such as cervical cancer screening in under 25s \cite{ali_retrospective_2013}; and $n=4000$ reflecting a cohort from a large network of hospitals, such as the Kaiser Permanente Northern California cervical cancer screening cohort \cite{schiffman_cohort_2016}.

\subsection{Comparator method and performance metrics}
To assess the model, we calculated the bias as the mean difference between the true parameter value and the median posterior estimate, the mean squared error (MSE) as the mean squared difference between the true parameter value and the median posterior estimate, the coverage as the percentage of simulations where the the $2.5\%$ quantile and the $97.5\%$ quantile interval contains the true parameter value and the width of the $95\%$ credible interval as the difference between the $97.5\%$ quantile and the $2.5\%$ quantile. 

To compare to a previous PI model, we used the `logistic-weibull' model in the PIMixture R package \cite{cheung_pimixture_2023}. This frequentist method gives a point estimate, standard error and confidence interval for each parameter in the model. Bootstrapping was used to calculate confidence intervals for survival probabilities and prevalence estimates. To summarise the performance of this model, we calculated the bias as the mean difference between the true parameter value and the point estimate; the MSE as the mean squared difference between the true parameter value and the point estimate; the coverage as the percentage of simulations where the the $2.5\%$ quantile and the $97.5\%$ quantile interval contains the true parameter value; and the width of the $95\%$ confidence interval as the difference between the upper and lower confidence interval. 

\subsection{Prior specification}
To reflect the scenario where we wish to specify priors using information from a previous study with a similar but not identical population to our dataset, we used the estimated parameters from those aged 45+ \cite{hyun_sampleweighted_2020} to create informative priors. The median event time in this age group was 23 months, so we chose a corresponding prior with 95\% quantiles of 15.5 months and 34 months. We assumed the event rate increases over time with a median value of 2 and 95\% quantiles 1.64 and 2.43. We used the confidence intervals for incidence, prevalence and cure probabilities to calculate the ratios and confidence intervals of prevalent to incidence and prevalent to cure. For strain 16, we derived the prior parameters using Equation \ref{eq: mu elicit} and \ref{eq: sigma elicit}. For strain 18, we compared the ratios with strain 16 to get the priors for $\beta_{2 \pi}$ and $\beta_{2 \delta}$. Strain 18 is known to cause cancer less quickly than strain 16, so we set a prior on the hazard ratio as 0.5, meaning $\gamma=-0.69$.
\begin{equation}
    \begin{split}
        \alpha & \sim \text{LogNormal}(\log(2), 0.1^2)\\
        \widetilde{m} & \sim \text{LogNormal}(\log(23/12), 0.2^2) \\
        \beta_{1\pi}& \sim  \text{Normal}(-0.07, 0.31^2)\\
        \beta_{2\pi}& \sim  \text{Normal}(-0.89, 0.33^2)\\
        \beta_{1\delta}& \sim  \text{Normal}(1.69, 0.22^2)\\
        \beta_{2\delta} & \sim  \text{Normal}(0.00, 0.05^2)\\
        \gamma & \sim  \text{Normal}(-0.69, 0.1^2)\\ 
    \end{split}
    \label{eq: priors cerv informative}
\end{equation}
To reflect a scenario where we have little to no prior information and wish to use vague priors, we set the multinomial regression and Weibull regression parameters to have mean 0, Weibull shape to have mean 1 and then set the Weibull mean event time to have a large variance:
\begin{equation}
    \begin{split}
        \alpha & \sim \text{LogNormal}(\log(1), 1^2)\\
        \widetilde{m} & \sim \text{LogNormal}(\log(2), 1^2) \\
        \beta_{1\pi}& \sim  \text{Normal}(0,1^2)\\
        \beta_{2\pi}& \sim  \text{Normal}(0,1^2)\\
        \beta_{1\delta}& \sim  \text{Normal}(0,1^2)\\
        \beta_{2\delta} & \sim  \text{Normal}(0,1^2)\\
        \gamma & \sim  \text{Normal}(0,1^2)\\ 
    \end{split}
    \label{eq: priors cerv vague}
\end{equation}
To reflect a scenario where the priors are misspecified, we reduce the variance of the vague priors:
\begin{equation}
    \begin{split}
        \alpha & \sim \text{LogNormal}(\log(1), 0.31^2)\\
        \widetilde{m} & \sim \text{LogNormal}(\log(2), 0.03^2) \\
        \beta_{1\pi}& \sim  \text{Normal}(0, 0.31^2)\\
        \beta_{2\pi}& \sim  \text{Normal}(0,0.31^2)\\
        \beta_{1\delta}& \sim  \text{Normal}(0,0.31^2)\\
        \beta_{2\delta} & \sim  \text{Normal}(0, 0.31^2)\\
        \gamma & \sim  \text{Normal}(0,0.26^2)\\ 
    \end{split}
    \label{eq: priors cerv misspec}
\end{equation}

\subsection{Simulation results}

We first compared the performance of our method under different choices for the prior distribution.
When $n=40$, as seen in Table \ref{tab: n40main}, the model with informative priors performs well and has the smaller mean bias than the vague prior for all parameters. For most parameters, coverage is better under the vague prior. However, the mean 95\% credible width is larger under the vague prior, meaning improved coverage comes at the cost of larger posterior uncertainty.

\begin{table}[ht]
\centering
\begin{tabular}{lrlrrrrr}
  \toprule
  & True & Prior & Mean & Bias & MSE & Cov. \% & CIW.  \\ \midrule
$\alpha$ & 2 & I  & 2.056 & \textbf{0.056} & \textbf{0.037} & \textbf{100.0} & \textbf{2.523}\\

   & & V & 1.115 & -0.887 & 1.123 & 98.6 & 6.321 \\ 
  $\lambda$ & 2.2 & I  & 2.272 & \textbf{0.072} & \textbf{0.034} & \textbf{100.0} & \textbf{1.746}  \\
   &  & V & 4.379 & 2.179 & 9.995 & \textbf{100.0} & 85.987 \\ 
  $\beta_{1\pi}$ & 0.799 & I& 0.027 & \textbf{-0.772} & \textbf{0.615} & 3.4 & \textbf{1.051} \\  
  &  & V & -0.128 & -0.927 & 1.002 & \textbf{69.8} & 2.274 \\ 
  $\beta_{1\delta}$ & 2.296 & I & 1.701 & \textbf{-0.595} & \textbf{0.360} & 0.2 & \textbf{0.763} \\ 
   & & V & 1.406 & -0.890 & 0.865 & \textbf{66.4} & 2.281\\ 
  $\beta_{2\pi}$ & -1.498 & I& -0.877 & \textbf{0.621} & \textbf{0.392} & 55.2 & \textbf{1.257} \\ 
  & & V & -0.355 & 1.143 & 1.445 & \textbf{85.8} & 3.215\\ 
  $\beta_{2\delta}$ & -0.762 & I& -0.001 & \textbf{0.761} & \textbf{0.580} & 0.0 & \textbf{0.196} \\  
   &  & V  & 0.214 & 0.976 & 1.158 & \textbf{91.8} & 3.04\\ 
  $\gamma$ & -0.5 & I & -0.689 & \textbf{-0.189} & \textbf{0.036} & 85 & \textbf{0.391}\\  
  & & V& -0.046 & 0.454 & 0.276 & \textbf{100.0} & 3.612 \\ 
   \hline
\end{tabular}
\caption{For sample size $n=40$, the true values and estimated mean, bias, mean squared error (MSE), coverage and 95\% coverage interval width (CIW) from 500 simulations for the Prevalence Incidence Cure model. Results are shown using an informative prior (I) and using a vague prior (V). Full results, including under a misspecified prior, are in Table \ref{tab: n40} in the Appendix.}
    \label{tab: n40main}
\end{table}

When $n=400$, as seen in Table \ref{tab: n400main}, the vague prior now has the smallest bias for all parameters. This is due to the informative prior not being centred on the true parameter value because the prior was based on a slightly different cohort to the data generation cohort (age group 22-44 vs 45+). This slightly misspecified prior, coupled with the small prior variance means the informative prior has a strong influence even at a sample size of $n=400$. However, the vague prior still has the larger 95\% credible interval widths than under the informative prior.
This suggests that an informative prior may be preferable at small sample sizes, but a vague prior might be preferred at larger sample sizes. As shown in Appendix Tables \ref{tab: n4000}, \ref{Tab: n4000Survproba}, \ref{Tab: n4000Survprobb}, as expected, as the sample size increases to 4000, the choice of prior becomes less influential and all results are similar. Additionally, as shown in Appendix Tables \ref{tab: n40} to \ref{Tab: n4000Survprobb}, the misspecified prior performs the worst of the three priors. \\

\begin{table}[ht]
\centering
\begin{tabular}{lrlrrrrr}
  \toprule
 & True & Prior & Mean  & Bias & MSE & Cov. \% & CIW.  \\ \midrule
$\alpha$ & 2 & I  & 2.123 & 0.123 & \textbf{0.114} & \textbf{98.2 }& \textbf{1.577}  \\
   & & V  & 2.000 & \textbf{$<$ 0.001} & 0.335 & 95.2 & 2.225 \\ 
  $\lambda$ & 2.2 & I  & 2.157 & -0.043 & \textbf{0.041} & \textbf{98.2} & \textbf{0.942}\\
   &  & V & 2.214 & \textbf{0.014} & 0.137 & 96.2 & 3.611 \\ 
  $\beta_{1\pi}$ & 0.799 & I & 0.316 & -0.483 & 0.248 & 9.0 & \textbf{0.655}   \\  
  &  & V   & 0.604 & \textbf{-0.195} & \textbf{0.082} & \textbf{90.6} & 0.985 \\ 
  $\beta_{1\delta}$ & 2.296 & I & 1.857 & -0.439 & 0.200 & 2.2 & \textbf{0.513} \\ 
   & & V & 2.127 & \textbf{-0.169} & \textbf{0.058} & \textbf{90.8} & 0.879 \\ 
  $\beta_{2\pi}$ & -1.498& I & -0.816 & 0.682 & 0.493 & 18.6 & \textbf{1.023} \\ 
  & & V  & -0.996 & \textbf{0.502} & \textbf{0.421} & \textbf{84.0} & 1.873  \\ 
  $\beta_{2\delta}$ & -0.762 & I& -0.008 & 0.754 & 0.569 & 0.0 & \textbf{0.194} \\ 
   &  & V & -0.372 & \textbf{0.390} & \textbf{0.260} & \textbf{85.8} & 1.463 \\ 
  $\gamma$ & -0.5 & I& -0.678 & -0.178 & \textbf{0.032} & 79.2 & \textbf{0.384 }\\ 
  & & V  & -0.335 & \textbf{0.165} & 0.197 & \textbf{97.8} & 1.951 \\ 
   \hline
\end{tabular}
\caption{For sample size $n=400$, the true values and estimated mean, bias, mean squared error (MSE), coverage and 95\% coverage interval width (CIW) from 500  simulations for the Prevalence Incidence Cure model. Results are shown using an informative prior (I) and using a vague prior (V). Full results, including under a misspecified prior, are in Table \ref{tab: n400} in the Appendix.}
    \label{tab: n400main}
\end{table}

Tables \ref{Tab: n40Survprobamain} and \ref{Tab: n400Survprobamain} compare estimates of survival (at 2 and 10 years), prevalence and cure under our PIC model and under the PI model.
When $n=40$, under the frequentist PI model (using the default settings), 26 out of 500 simulations failed to converge and so we removed these simulations when calculating performance metrics for the PI model. In a further 46 simulations, the model failed to calculate confidence intervals: we retained these simulations when calculating performance metrics for point estimates, but removed them when calculating performance metrics for confidence intervals. For sample sizes $n=400$ and $n=4000$, the PI model converged in all 500 simulations.\\
\begin{table}[ht]
\centering
\begin{tabular}{p{1.3cm}ccrrrrr}
    \toprule
        & True & Model & Mean & Bias & MSE & Cov. \% & CIW.  \\ \midrule
\multirow{3}{\linewidth}{2 year survival}  &\multirow{3}{*}{0.788} & I & 0.788 & \textbf{$<$ 0.001} & \textbf{0.001} & \textbf{99.8} & \textbf{0.163} \\ 
  &  &   V & 0.762 & -0.026 & 0.005 & 96.2 & 0.276 \\ 
   &  & PI & 0.793 & 0.005 & 0.005 & 97.0 & 0.299 \\ 
  \cmidrule{1-2}
  \multirow{3}{\linewidth}{10 year survival} &\multirow{3}{*}{0.755} & I & 0.728 & \textbf{-0.027} & \textbf{0.001} & \textbf{97.8} & \textbf{0.167} \\  
  &  &   V & 0.701 & -0.054 & 0.008 & 91.4 & 0.299 \\ 
  &  &  PI & 0.688 & -0.100 & 0.020 & 83.9 & 0.346 \\ 
  \cmidrule{1-2}
    \multirow{3}{\linewidth}{Preval\-ence}  & \multirow{3}{*}{0.169} & I & 0.138 & -0.031 & \textbf{0.001} & \textbf{97.4} & \textbf{0.134} \\ 
  &  &  V& 0.156 & \textbf{-0.013} & 0.003 & 96.6 & 0.223 \\ 
   &  &PI& 0.190 & 0.021 & 0.006 & 93.5 & 0.298 \\ 
  \cmidrule{1-2}
  \multirow{2}{\linewidth}{Cure} & \multirow{2}{*}{0.755} & I & 0.727 & \textbf{ -0.028} & \textbf{0.001} &\textbf{ 97.8} & \textbf{0.167} \\ 
  &  & V & 0.668 & -0.087 & 0.012 & 86.0 & 0.387 \\ 
\bottomrule
\end{tabular}\caption{For sample size $n=40$, the true values and estimated mean, bias, mean squared error (MSE), coverage (Cov) and 95\% coverage interval width (CIW) for survival probability at times $t=2$ and $t=10$, prevalence and cure under strain 16. The results compare the PI model, and the Prevalence Incidence Cure model using an informative prior (I) and vague prior (V). Full results for strain 18 and under a misspecified prior are in Table \ref{Tab: n40Survproba} in the Appendix.}
    \label{Tab: n40Survprobamain}
\end{table}

\begin{table}[ht]
\centering
\begin{tabular}{p{1.3cm}ccrrrrr}
    \toprule
        Estimand  & True & Model & Mean & Bias & MSE & Cov. \% & CIW.  \\ \midrule

\multirow{3}{\linewidth}{2 year survival}  &\multirow{3}{*}{0.788} & I & 0.777 & -0.011 & \textbf{0.001} & 93.8 & \textbf{0.083} \\ 
  &  & V & 0.785 & \textbf{-0.003} & 0.001 & \textbf{95.2} & 0.094 \\ 
   &  & PI & 0.794 & 0.006 & 0.001 & 94.8 & 0.093  \\ 
  \cmidrule{1-2}
  \multirow{3}{\linewidth}{10 year survival}  &\multirow{3}{*}{0.755} & I& 0.728 & -0.027 & 0.001 & 77.4 & \textbf{0.085}   \\ 
  &   & V & 0.746 & \textbf{-0.009} & \textbf{0.001} & \textbf{96.2} & 0.100  \\ 
  &   & PI& 0.689 & -0.100 & 0.011 & 31.2 & 0.110\\ 
  \cmidrule{1-2}
    \multirow{3}{*}{Prev.}  & \multirow{3}{*}{0.169} & I & 0.157 & -0.012 & \textbf{$<$ 0.001} & 91.4 & \textbf{0.072}   \\ 
   &  & V & 0.164 & \textbf{-0.005} & $<$ 0.001 & \textbf{94.6} & 0.082\\ 
   &   &PI & 0.193 & 0.024 & 0.001 & 82.6 & 0.094   \\
  \cmidrule{1-2}
  \multirow{2}{*}{Cure}  & \multirow{2}{*}{0.755} & I  & 0.728 & -0.027 & \textbf{0.001} & 77.4 & \textbf{0.085}    \\ 
  & &  V & 0.745 & \textbf{-0.010} & 0.001 & \textbf{96.0} & 0.103\\ 
   \hline
\end{tabular}\caption{For sample size $n=400$, the true values and estimated mean, bias, mean squared error (MSE), coverage (Cov) and 95\% coverage interval width (CIW) for survival probability at time $t=2$ and $t=10$ and estimates for cure and prevalence under strain 16. Models compared are PI (Prevalence Incidence), I a Prevalence Incidence Cure model using informative priors and V a Prevalence Incidence Cure model using vague priors. Full results for strain 18 and misspecified priors can be seen in Table \ref{Tab: n400Survproba} in the Appendix. }
    \label{Tab: n400Survprobamain}
\end{table}

Tables \ref{Tab: n40Survprobamain} and  \ref{Tab: n400Survprobamain} show the PI model slightly over-estimates the survival probability at 2 years. The PIC models underestimate the survival probabilities. The informative prior has less bias than the PI model at 2 years with $n=40$ but greater bias with $n=400$. This is reversed for the vague prior which is more biased at 2 years with $n=40$ and less biased with $n=400$. The PI model under-estimates the survival probability at 10 years whilst the PIC models with vague and informative priors have smaller bias at 10 years. The under-estimation of the PI model at 10 years is due to the PI survival function assuming a plateau at 0 as time goes to infinity, meaning survival probabilities at later time points will be smaller than the PIC model, which plateaus at the cure probability.

The PI model also overestimates prevalence whilst the PIC models with vague and informative priors have less bias. The PI model makes the assumption that the survival probability tends to 0 as time goes to infinity which means the incidence distribution can be biased if this is untrue, as it is in the simulation study. It would appear in this situation the incidence distribution underestimates the probability of events at small timepoints, seen also by the overestimation of 2 year survival, which means missing baseline results are more likely to be estimated as a prevalent case than an incidence case. This leads to an overestimation of prevalence.

The coverage of the survival probabilities decreases as the sample size increases for the PI model. This can be explained by the PI model becoming more confident due to the larger sample size, but concentrating around incorrect values. Results for strain 18 are shown in the Appendix with similar patterns shown: strain 18 is less prevalent, so the confidence intervals and credible intervals are larger.

The PIC models using vague and informative priors estimate the cure probability well.

\section{Application to Diabetic Macular Oedema}
We apply our method to a dataset from Moorfields Eye Hospital, UK, of 2614 eyes undergoing treatment for DMO over 48 months. The data are openly available in the `eyedata' package in R \cite{heeren_eyedata_2020}. A full description of the cohort can be found in Kern et al \cite{kern_open-source_2021}. The event of interest is the time after the initiation of treatment until `good' eyesight, defined as a VA value of 70 or greater, is first achieved.

\begin{figure}
    \centering
    \includegraphics[width=1\linewidth]{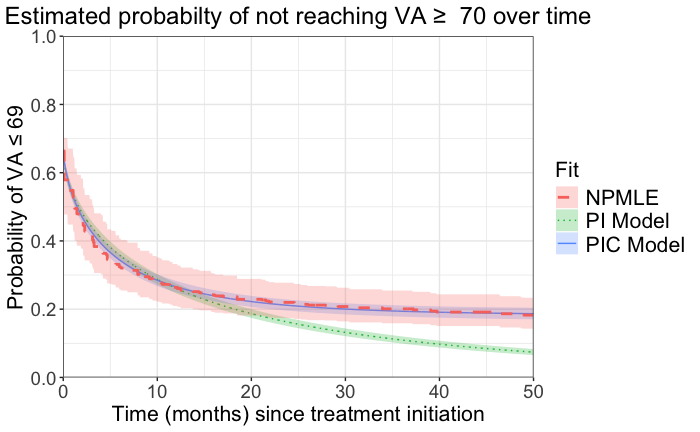}
    \caption{Time to a VA value of 70 or above after starting treatment in the DMO application under three different models. The shaded areas represent the 95\% credible interval for the PIC model and the 95\% confidence interval for the PI and NPMLE model.}
    \label{fig:DMO data}
\end{figure}

Figure \ref{fig:DMO data} shows a non-parametric maximum likelihood estimate (NPMLE) of the data, using the Turnbull algorithm \cite{turnbull_nonparametric_1974, therneau_survival_2001}. There is a clear vertical line at time 0, which is indicative of prevalence, and a plateau at the end, which is indicative of a cure proportion. A previous analysis \cite{kern_open-source_2021} excluded patients who have no recorded VA value at treatment initiation and treated prevalent cases as incident cases with event time zero. The abstract reported estimate from this analysis is the time until $50\%$ of eyes achieve `good' eyesight, despite ~30\% of the population being prevalent at baseline. This analysis approach inflates the estimates of the benefit of treatment and makes it hard to compare cohorts who have different proportions of `good' eyesight at treatment initiation. It is also more relevant to the patient to provide their prognosis conditional on their baseline VA measurement. Furthermore, it does not provide an estimate for the proportion of patients who never reach a value of 70. 

There were 6 eyes with a missing VA value at baseline, meaning disease prevalence could not be ascertained for them. Of eyes with a valid baseline VA value, 884 out of 2608 were above 70. We apriori assume the median time to a VA value of $\geq 70$ is 10 months with 95\% quantiles of 4.2 and 24.0 months, since few improvements in VA  are seen after 12 months \cite{fong_treatment_2018}.
\begin{equation*}
            \widetilde{m} \sim \text{LogNormal}(\log(10), 0.20)\\
\end{equation*}
We assume apriori that the event rate is approximately constant, with $\alpha$ around 1, with apriori median of 1 and 95\% quantiles of 0.25 and 4.
\begin{equation*}
        \alpha \sim \text{LogNormal}(\log(1), 0.5)
\end{equation*}
Finally, we assume apriori that the ratios for prevalence to incidence and cure to incidence have 95\% confidence intervals of (0.05, 0.5) \cite{maggio_antivascular_2018, fong_treatment_2018}. Using Eq. \ref{eq: ratio pi} and \ref{eq: ratio delta} we get:
\begin{equation}
    \begin{split}
        \beta_{1\pi}& \sim \text{Normal}(-1.84, 0.41^2)\\
        \beta_{1\delta}& \sim \text{Normal}(-1.84, 0.41^2)\\
    \end{split}
\end{equation}
Figure \ref{fig:DMO data} shows the observed data using the Turnbull algorithm, the fitted PIC model and the fitted model from PIMixture. Figure \ref{fig:DMO data} visually shows a substantially better fit to the non parametric estimate when using the PIC model compared to the PI model as well as a decrease in 95\% credible/ confidence interval between the non parametric and PIC model. In particular, at 48 months the survival probability is 8\% under the PI model compared to 18\% under the PIC model. This large difference suggests the PIC model can provide meaningfully substantively different answers to the PI model. This echoes the simulation study, which shows the same effect of PI overestimating survival in early event times and then underestimating survival in later event times.

Posterior parameter estimates are given in Table \ref{tab:DMO2}. Both models give similar estimates, of around a third, for prevalence (patients receiving treatment who already have `good' eyesight). In the original analysis by Kern et al, the median time to VA greater than 70 including prevalence cases was reported as 1.9 months, whilst using the PIC estimates gives 1.50 months and PI estimates gives 1.59 months. This decrease could be due to the PI and PIC models including patients who had missing results at baseline, whom the original analysis dropped. Additionally, using the PIC model estimates, the median time to VA greater than 70, given not prevalence at baseline, is 6.44 months and using the PI model estimates is 7.32 months. This means, upon treatment initiation when patients should have their VA measured, they can be informed of their prognosis given their baseline measurement. The PIC model further estimates around 18\% of patients will be non responders.

\begin{table}[ht!]
\centering
\begin{tabular}{l|ll}
  \toprule
 & PIC & PI\\ 
  \midrule
  $\alpha$ & 0.68 (0.64, 0.72)  &0.60 (0.58, 0.63)\\ 
  $\lambda$ & 5.27 (4.71, 5.97) & 
13.48 (12.36, 14.70)\\ 
  Prevalence & 0.33 (0.31, 0.35) &  0.34 (0.32, 0.36)\\ 
  Cure & 0.18 (0.16, 0.20) & \\
   \bottomrule
\end{tabular}
\caption{Estimates for models applied to the DMO application. Median estimate (95\% credible interval) under the PIC model and point estimate (95\% confidence interval) under the PI model.}
\label{tab:DMO2}
\end{table}

\section{Discussion}

We have proposed a Prevalence-Incidence-Cure (PIC) model, which extends previous work on Prevalence Incidence models by explicitly incorporating a cure proportion, which can be influenced by covariates. Our approach provides more accurate estimates of prevalence and incidence times in settings involving a cure fraction and with incomplete baseline prevalence data.

Consideration of priors was made and as seen in the simulation study, using an informative prior can reduce bias, especially at smaller sample sizes. However, as the sample size increases, the risk of incorrectly specifying an informative prior may outweigh the benefits, and a vague prior may be preferable.

The results from the analysis of DMO supports existing evidence that approximately 15\% of patients are non-responders to VEGF \cite{krebs_non-responders_2013}. Inclusion of further covariates can help identify which patients are at risk of being non-responders. This model does not provide insight as to how long patients remain with a VA value of above 70 and evaluates a binary event of a value measured on a continuous scale. It furthermore does not take into consideration the correlation between 2 eyes in a patient and treats each eye as independent. Therefore, for full understanding of the long term prognosis for patients treated with VEGF further analysis and model development is required.

Our PIC model may offer only limited benefit over existing PI models if the cure proportion is very small, or if the follow up time is not long enough to observe the cure plateau. Furthermore, if there isn't any missing diagnoses at baseline the model just collapses into a cure model. However, there are several further examples where missing baseline results and a sizeable cure proportion are present, particularly surrounding population wide screening services such as for breast cancer, cervical cancer and diabetic eye screening.  Other applications could include secondary care hospital settings where regular observation checks are performed but data on symptoms at admissions may be incomplete, such as recording of stroke symptoms via the NIH stroke score (NIHSS) \cite{reeves_variation_2015}. 

There are several extensions to our approach that could be considered in future work. We have not considered the sensitivity and specificity of diagnostic test results, but we anticipate an extension based on the previous proposals for PI models \cite{klausch_bayesian_2025} would be feasible. Extensions to other parametric and non-parametric survival-time distributions, such as Cox proportional hazards, would allow for further flexibility. This extension would require only the survival function to be replaced in the JAGS code. For example, a Bayesian Cox proportional hazards model could be implemented assuming a piecewise hazard function \cite{lunn_bugs_2012}. Finally, extending our approach to allow for time-varying covariates would allow for our approach to be adopted for dynamic risk prediction \cite{kurtz_dynamic_2019}.

\clearpage

\newpage



\clearpage
\section{Appendix}
\subsection{Full simulation results tables}

\begin{table}[ht]
\centering
\begin{tabular}{rrrrrrrr}
  \toprule
 Para. & True & Priors & Mean & Bias & MSE & Cov. \% & CIW.  \\ \midrule
$\alpha$ & 2 & I  & 2.056 & \textbf{0.056} & \textbf{0.037} & \textbf{100.0} & 2.523\\
   & & M & 0.943 & -1.057 & 1.130 & 16.2 & \textbf{1.327} \\ 
   & & V & 1.115 & -0.887 & 1.123 & 98.6 & 6.321 \\ 
  $\lambda$ & 2.2 & I  & 2.272 & \textbf{0.072} & \textbf{0.034} & \textbf{100.0} & \textbf{1.746}  \\
  &  & M  & 2.976 & 0.776 & 0.621 & 0.0 & 1.891 \\ 
   &  & V & 4.379 & 2.179 & 9.995 & \textbf{100.0} & 85.987 \\ 
  $\beta_{1\pi}$ & 0.799 & I& 0.027 & \textbf{-0.772} & \textbf{0.615} & 3.4 & 1.051 \\ 
  &  & M & -0.239 & -1.038 & 1.091 & 0.0 & \textbf{1.012} \\ 
  &  & V & -0.128 & -0.927 & 1.002 & \textbf{69.8} & 2.274 \\ 
  $\beta_{1\delta}$ & 2.296 & I & 1.701 & \textbf{-0.595} & \textbf{0.360} & 0.2 & \textbf{0.763} \\ 
   & & M & 0.663 & -1.633 & 2.680 & 0.0 & 0.993  \\ 
   & & V & 1.406 & -0.890 & 0.865 & \textbf{66.4} & 2.281\\ 
  $\beta_{2\pi}$ & -1.498 & I& -0.877 & 0.621 & \textbf{0.392} & 55.2 & 1.257 \\ 
   &  & M  & -0.101 & 1.397 & 1.957 & 0.0 & \textbf{1.175}\\ 
  & & V &-0.355 & 1.143 & 1.445 & \textbf{85.8} & 3.215\\ 
  $\beta_{2\delta}$ & -0.762 & I& -0.001 & \textbf{0.761} & \textbf{0.580} & 0.0 & \textbf{0.196} \\ 
   &  & M & 0.152 & 0.914 & 0.844 & 0.2 & 1.159 \\ 
   &  & V  & 0.214 & 0.976 & 1.158 & \textbf{91.8} & 3.040\\ 
  $\gamma$ & -0.5 & I & -0.689 & \textbf{-0.189} & \textbf{0.036} & 85.0 & \textbf{0.391}\\ 
   &  & M & -0.018 & 0.482 & 0.234 & 74.0 & 0.999  \\ 
  & & V& -0.046 & 0.454 & 0.276 & \textbf{100.0} & 3.612 \\ 
   \hline
\end{tabular}
\caption{For sample size $n=40$, the true values and estimated mean, bias, mean squared error (MSE), coverage (Cov) and 95\% coverage interval width (CIW) for estimates from simulations comparing I, a Prevalence Incidence Cure model using informative priors, V, a Prevalence Incidence Cure model using vague priors and M, a Prevalence Incidence Cure model using Misspecified priors.}
    \label{tab: n40}
\end{table}

\begin{table}[ht]
\centering
\begin{tabular}{p{1.3cm}cccrrrrr}
    \toprule
         & S. & True & Model & Mean & Bias & MSE & Cov. \% & CIW.  \\ \midrule
\multirow{8}{\linewidth}{2 year survival}  &\multirow{4}{*}{16} &\multirow{4}{*}{0.788} & I & 0.788 & \textbf{$<$ 0.001} & \textbf{0.001} & \textbf{99.8} & \textbf{0.163} \\ 
   &  & & M  & 0.653 & -0.135 & 0.019 & 4.0 & 0.194 \\ 
  &  &  & V & 0.762 & -0.026 & 0.005 & 96.2 & 0.276 \\ 
   &  & & PI & 0.793 & 0.005 & 0.005 & 97.0 & 0.299 \\ 
   \cmidrule{2-2}
   & \multirow{4}{*}{18} & \multirow{4}{*}{0.855} & I& 0.887 & \textbf{0.032} & \textbf{0.001} & 97.8 & \textbf{0.128}  \\ 
  &  & & M & 0.693 & -0.162 & 0.028 & 8.4 & 0.283 \\ 
  &  & & V & 0.819 & -0.036 & 0.007 & \textbf{98.4} & 0.38 \\ 
  &  & & PI  & 0.888 & -0.034 & 0.012 & 97.8 & 0.73 \\ 
  \cmidrule{1-2}
  \multirow{8}{\linewidth}{5 year survival}   &\multirow{4}{*}{16} &\multirow{4}{*}{0.755} & I & 0.731 & -0.024 & \textbf{0.001} & \textbf{98.4} & \textbf{0.167}  \\ 
  &  & & M & 0.570 & -0.185 & 0.036 & 0.2 & 0.227 \\ 
  &  & &V & 0.719 & -0.036 & 0.006 & 94.8 & 0.291 \\ 
  &  & & PI & 0.758 & \textbf{0.002} & 0.007 & 95.6 & 0.31 \\ 
  \cmidrule{2-2}
   & \multirow{4}{*}{18} &\multirow{4}{*}{ 0.763} & I & 0.807 & 0.044 & \textbf{0.002} & 96.4 & \textbf{0.153}\\ 
  &  & &  M & 0.616 & -0.147 & 0.024 & 57.4 & 0.321 \\ 
  &  & & V & 0.777 & \textbf{0.014} & 0.009 & \textbf{98.6} & 0.417  \\ 
  &  & &  PI & 0.812 & -0.049 & 0.022 & 96.7 & 0.741 \\ 
  \cmidrule{1-2}
  \multirow{8}{\linewidth}{10 year survival}  &\multirow{4}{*}{16} &\multirow{4}{*}{0.755} & I & 0.728 & \textbf{-0.027} & \textbf{0.001} & \textbf{97.8} & \textbf{0.167} \\ 
  &  & & M  & 0.536 & -0.219 & 0.049 & 0.0 & 0.234 \\ 
  &  &  & V & 0.701 & -0.054 & 0.008 & 91.4 & 0.299 \\ 
  &  & & PI & 0.688 & -0.100 & 0.020 & 83.9 & 0.346 \\ 
  \cmidrule{2-2}
  & \multirow{4}{*}{18} & \multirow{4}{*}{0.756} & I& 0.792 & 0.036 & \textbf{0.002} & \textbf{98.6} & \textbf{0.146}  \\ 
  &  & & M & 0.583 & -0.173 & 0.032 & 36.2 & 0.333  \\ 
  &  & & V  & 0.760 & \textbf{0.004} & 0.009 & 98.6 & 0.435 \\ 
 &  & & PI & 0.687 & -0.069 & 0.039 & 87.6 & 0.75 \\ 
\end{tabular}\caption{For sample size $n=40$, the true values and estimated mean, bias, mean squared error (MSE), coverage (Cov) and 95\% coverage interval width (CIW) for survival probability at time $t=2$, $t=5$ and $t=10$ under strain 16 and strain 18. Models compared are PI (Prevalence Incidence), I a Prevalence Incidence Cure model using informative priors, V a Prevalence Incidence Cure model using vague priors and M a Prevalence Incidence Cure model using Misspecified priors.}
    \label{Tab: n40Survproba}
\end{table}
\begin{table}[ht]
\centering
\begin{tabular}{cc|rrrrrrr}
    \toprule
        & S. & True & Model & Mean & Bias & MSE & Cov. \% & CIW.  \\ \midrule
  \multirow{8}{*}{Prev.} & \multirow{4}{*}{16} & \multirow{4}{*}{0.169} & I & 0.138 & -0.031 & \textbf{0.001} & \textbf{97.4} & \textbf{0.134} \\ 
  &  & & M & 0.212 & 0.043 & 0.003 & 93.6 & 0.175\\ 
   &  & & V& 0.156 & \textbf{-0.013} & 0.003 & 96.6 & 0.223 \\ 
   &  & &PI& 0.190 & 0.021 & 0.006 & 93.5 & 0.298 \\ 
   \cmidrule{2-2}
   & \multirow{4}{*}{18}& \multirow{4}{*}{0.081} & I& 0.063 & -0.018 & \textbf{$<$0.001} & 100.0 & \textbf{0.103} \\ 
  &  & & M& 0.180 & 0.098 & 0.011 & 40.4 & 0.236\\ 
  &  & &V & 0.103 & 0.022 & 0.004 & \textbf{98.6} & 0.298 \\ 
  &  & &PI & 0.087 & \textbf{0.006} & 0.012 & 92.8 & 0.740 \\ 
  \cmidrule{1-2}
  \multirow{6}{*}{Cure} & \multirow{3}{*}{16} & \multirow{3}{*}{0.755} & I & 0.727 & \textbf{-0.028} & \textbf{0.001} & \textbf{97.8} & \textbf{0.167} \\ 
  &  & & M & 0.518 & -0.237 & 0.057 & 0.0 & 0.240  \\ 
  &  & & V & 0.668 & -0.087 & 0.012 & 86.0 & 0.387 \\ 
  \cmidrule{2-2}
  &  \multirow{3}{*}{18} & \multirow{3}{*}{0.756} & I & 0.79 & 0.034 & \textbf{0.002} & \textbf{98.8} & \textbf{0.146} \\ 
  &  & & M & 0.564 & -0.192 & 0.039 & 21.2 & 0.343 \\ 
  &  & & V & 0.723 & \textbf{-0.033} & 0.011 & 98.2 & 0.555\\ 
   \hline
\end{tabular}\caption{For sample size $n=40$, the true values and estimated mean, bias, mean squared error (MSE), coverage (Cov) and 95\% coverage interval width (CIW) for cure and prevalence under strain 16 and strain 18. Models compared are PI (Prevalence Incidence), PIC (Prevalence Incidence Cure) with I using informative priors, V vague priors and M Misspecified priors.}
    \label{Tab: n40Survprobb}
\end{table}

\begin{table}[ht]
\centering
\begin{tabular}{rlllllll}
  \toprule
 Para.& True & Priors & Mean  & Bias & MSE & Cov. \% & CIW.  \\ \midrule
$\alpha$ & 2.000 & I  & 2.123 & 0.123 & \textbf{0.114} & \textbf{98.2} & 1.577  \\
   &  & M & 1.449 & -0.551 & 0.380 & 72.6 & \textbf{1.407}  \\ 
   &  & V & 2.000 & \textbf{0.000} & 0.335 & 95.2 & 2.225 \\ 
$\lambda$ & 2.200 & I  & 2.157 & -0.043 & \textbf{0.041} & \textbf{98.2} & \textbf{0.942} \\
   &  & M & 2.613 & 0.413 & 0.205 & 2.0 & 1.049 \\ 
   &  & V & 2.214 & \textbf{0.014} & 0.137 & 96.2 & 3.611 \\ 
$\beta_{1\pi}$ & 0.799 & I & 0.316 & -0.483 & 0.248 & 9.0 & 0.655   \\ 
   &  & M & 0.042 & -0.757 & 0.586 & 0.0 & 0.665 \\ 
   &  & V & 0.604 & \textbf{-0.195} & \textbf{0.082} & \textbf{90.6} & \textbf{0.985} \\ 
$\beta_{1\delta}$ & 2.296 & I & 1.857 & -0.439 & 0.200 & 2.2 & 0.513 \\ 
   &  & M & 1.534 & -0.762 & 0.588 & 0.0 & \textbf{0.577} \\ 
   &  & V & 2.127 & \textbf{-0.169} & \textbf{0.058} & \textbf{90.8} & 0.879 \\ 
$\beta_{2\pi}$ & -1.498 & I & -0.816 & 0.682 & 0.493 & 18.6 & \textbf{1.023} \\ 
   &  & M & -0.269 & 1.229 & 1.527 & 0.0 & 0.991 \\ 
   &  & V & -0.996 & \textbf{0.502} & \textbf{0.421} & \textbf{84.0} & 1.873  \\ 
$\beta_{2\delta}$ & -0.762 & I & -0.008 & 0.754 & 0.569 & 0.0 & \textbf{0.194} \\ 
   &  & M & 0.142 & 0.904 & 0.838 & 0.0 & 0.876 \\ 
   &  & V & -0.372 & \textbf{0.390} & \textbf{0.260} & \textbf{85.8} & 1.463 \\ 
$\gamma$ & -0.500 & I & -0.678 & \textbf{-0.178} & \textbf{0.032} & 79.2 & \textbf{0.384} \\ 
   &  & M & -0.013 & 0.487 & 0.246 & 26.2 & 0.860 \\ 
   &  & V & -0.335 & 0.165 & 0.197 & \textbf{97.8} & 1.951 \\ 
  \hline
\end{tabular}
\caption{For sample size $n=400$, the true values and estimated mean, bias, mean squared error (MSE), coverage (Cov) and 95\% coverage interval width (CIW) comparing I, a Prevalence Incidence Cure model using informative priors, V, a Prevalence Incidence Cure model using vague priors and M, a Prevalence Incidence Cure model using Misspecified priors.}
\label{tab: n400}
\end{table}

\begin{table}[ht]
\centering
\begin{tabular}{p{1.3cm}cccrrrrr}
    \toprule
        & S. & True & Model & Mean & Bias & MSE & Cov. \% & CIW.  \\ \midrule
\multirow{8}{\linewidth}{2 year survival}   
  &\multirow{4}{*}{16} &\multirow{4}{*}{0.788} & I & 0.777 & -0.011 & \textbf{0.001} & 93.8 & 0.083 \\ 
  &  & & M & 0.768 & -0.021 & 0.001 & 84.6 & \textbf{0.080}  \\ 
  &  &  & V & 0.785 & \textbf{-0.003} & 0.001 & \textbf{95.2} & 0.094 \\ 
  &  & & PI & 0.794 & 0.006 & 0.001 & 94.8 & 0.093  \\ 
  \cmidrule{2-2}
  & \multirow{4}{*}{18} & \multirow{4}{*}{0.855} & I & 0.877 & 0.022 & \textbf{0.001} & 88.4 & \textbf{0.082} \\ 
  &  & & M & 0.816 & -0.038 & 0.002 & 82.0 & 0.123   \\ 
  &  & & V & 0.843 & \textbf{-0.012} & 0.001 & \textbf{95.0} & 0.141  \\ 
  &  & & PI & 0.883 & -0.029 & 0.002 & 92.8 & 0.132 \\ 
  \cmidrule{1-2}
\multirow{8}{\linewidth}{5 year survival}   
  &\multirow{4}{*}{16} &\multirow{4}{*}{0.755} & I & 0.730 & -0.026 & 0.001 & 78.0 & \textbf{0.085}  \\ 
  &  & & M & 0.707 & -0.048 & 0.003 & 47.2 & 0.097 \\ 
  &  & & V & 0.749 & -0.007 & \textbf{0.001} & \textbf{96.2} & 0.099 \\ 
  &  & & PI & 0.761 & \textbf{0.005} & 0.001 & 94.4 & 0.097\\ 
  \cmidrule{2-2}
  & \multirow{4}{*}{18} &\multirow{4}{*}{0.763} & I & 0.805 & 0.042 & 0.002 & 63.0 & \textbf{0.091}  \\ 
  &  & & M & 0.759 & \textbf{-0.004} & \textbf{0.001} & \textbf{98.4} & 0.147\\ 
  &  & & V & 0.776 & 0.013 & 0.002 & 94.2 & 0.170 \\ 
  &  & & PI & 0.833 & -0.070 & 0.006 & 64.4 & 0.142\\ 
  \cmidrule{1-2}
\multirow{8}{\linewidth}{10 year survival}  
  &\multirow{4}{*}{16} &\multirow{4}{*}{0.755} & I & 0.728 & -0.027 & 0.001 & 77.4 & \textbf{0.085}   \\ 
  &  & & M & 0.696 & -0.060 & 0.004 & 26.6 & 0.098 \\ 
  &  & & V & 0.746 & \textbf{-0.009} & \textbf{0.001} & \textbf{96.2} & 0.100  \\ 
  &  & & PI & 0.689 & -0.100 & 0.011 & 31.2 & 0.110\\ 
  \cmidrule{2-2}
  & \multirow{4}{*}{18} & \multirow{4}{*}{0.756} & I & 0.796 & 0.040 & 0.002 & 64.4 & \textbf{0.090}  \\ 
  &  & & M & 0.747 & \textbf{-0.008} & \textbf{0.001} & \textbf{97.6} & 0.150  \\ 
  &  & & V & 0.767 & 0.011 & 0.002 & 94.0 & 0.175  \\ 
  &  & & PI & 0.728 & -0.028 & 0.003 & 89.0 & 0.183\\ 
\end{tabular}
\caption{For sample size $n=400$, the true values and estimated mean, bias, mean squared error (MSE), coverage (Cov) and 95\% coverage interval width (CIW) for survival probability at time $t=2$, $t=5$ and $t=10$ and estimates for cure and prevalence under strain 16 and strain 18. Models compared are PI (Prevalence Incidence), I a Prevalence Incidence Cure model using informative priors, V a Prevalence Incidence Cure model using vague priors, and M a Prevalence Incidence Cure model using Misspecified priors.}
\label{Tab: n400Survproba}
\end{table}

\begin{table}[ht]
\centering
\begin{tabular}{cc|ccccccccccccc}
    \toprule
        & S. & True & Model & Mean & Bias & MSE & Cov. \% & CIW.  \\ \midrule
  \multirow{8}{*}{Prev.} 
  & \multirow{4}{*}{16} & \multirow{4}{*}{0.169} & I & 0.157 & -0.012 & \textbf{$<$ 0.001} & 91.4 & \textbf{0.072}   \\ 
  &  & & M & 0.157 & -0.012 & $<$ 0.001 & 91.8 & 0.072 \\ 
  &  & & V & 0.164 & \textbf{-0.005} & $<$ 0.001 & \textbf{94.6} & 0.082\\ 
  &  & & PI & 0.193 & 0.024 & 0.001 & 82.6 & 0.094   \\ 
  \cmidrule{2-2}
  & \multirow{4}{*}{18} & \multirow{4}{*}{0.081} & I  & 0.077 & \textbf{-0.004} & \textbf{$<$0.001} & \textbf{98.4} & \textbf{0.075} \\ 
  &  & & M & 0.113 & 0.032 & 0.001 & 78.0 & 0.099  \\ 
  &  & & V & 0.094 & 0.013 & 0.001 & 94.4 & 0.113  \\ 
  &  & & PI & 0.097 & 0.016 & 0.001 & 89.2 & 0.134  \\ 
  \cmidrule{1-2}
  \multirow{6}{*}{Cure} 
  & \multirow{3}{*}{16} & \multirow{3}{*}{0.755} & I  & 0.728 & -0.027 & 0.001 & 77.4 & \textbf{0.085}    \\ 
  &  & & M & 0.693 & -0.062 & 0.004 & 24.0 & 0.100 \\ 
  &  & & V & 0.745 & \textbf{-0.010} & \textbf{0.001} & \textbf{96.0} & 0.103\\ 
  \cmidrule{2-2}
  & \multirow{3}{*}{18} & \multirow{3}{*}{0.756} & I & 0.796 & 0.040 & 0.002 & 65.2 & \textbf{0.090}\\ 
  &  & & M & 0.745 & -0.011 & \textbf{0.001} & \textbf{98.2} & 0.153   \\ 
  &  & & V  & 0.766 & \textbf{0.010} & 0.002 & 95.2 & 0.180  \\ 
   \hline
\end{tabular}
\caption{For sample size $n=400$, the true values and estimated mean, bias, mean squared error (MSE), coverage (Cov) and 95\% coverage interval width (CIW) for survival probability at time $t=2$, $t=5$ and $t=10$ and estimates for cure and prevalence under strain 16 and strain 18. Models compared are PI (Prevalence Incidence), PIC (Prevalence Incidence Cure) with I using informative priors, V vague priors and M Misspecified priors.}
\label{Tab: n400Survprobb}
\end{table}

\begin{table}[ht]
\centering
\begin{tabular}{rlllllll}
  \toprule
 Para. & True  & Priors & Mean  & Bias & MSE & Cov. \% & CIW. \\ \midrule
$\alpha$ & 2 & I & 1.993 & -0.007 & 0.018 & \textbf{95.8} & 0.551 \\
   & & M & 1.944 & \textbf{-0.056} & \textbf{0.017} & 95.6 & 0.523   \\ 
   & & V & 1.942 & -0.058 & 0.024 & 92.8 & \textbf{0.574}  \\ 
$\lambda$ & 2.2 & I & 2.091 & -0.109 & 0.021 & 79.8 & 0.373  \\
   & & M & 2.357 & 0.157 & 0.026 & 19.0 & \textbf{0.261} \\ 
   & & V & 2.132 & \textbf{-0.068} & \textbf{0.019} & \textbf{91.0} & 0.455 \\ 
$\beta_{1\pi}$ & 0.799 & I & 0.576 & -0.223 & 0.054 & 12.0 & \textbf{0.281} \\ 
   & & M & 0.589 & -0.210 & 0.048 & 19.8 & 0.300 \\ 
   & & V & 0.809 & \textbf{0.010} & \textbf{0.007} & \textbf{95.6} & 0.344\\ 
$\beta_{1\delta}$ & 2.296 & I & 2.071 & -0.225 & 0.054 & 2.6 & \textbf{0.232}\\ 
   & & M & 2.107 & -0.189 & 0.039 & 18.6 & 0.263 \\ 
   & & V & 2.312 & \textbf{0.016} & \textbf{0.006} & \textbf{95.4} & 0.306\\ 
$\beta_{2\pi}$ & -1.498 & I & -0.847 & 0.651 & 0.438 & 0.2 & \textbf{0.497} \\ 
   & & M & -0.970 & 0.528 & 0.295 & 2.4 & 0.573  \\ 
   & & V & -1.414 & \textbf{0.084} & \textbf{0.038} & \textbf{92.2} & 0.677 \\ 
$\beta_{2\delta}$ & -0.762 & I & -0.086 & 0.676 & 0.457 & 0.0 & \textbf{0.183}\\ 
   & & M & -0.387 & 0.375 & 0.151 & 5.0 & 0.446   \\ 
   & & V & -0.713 & \textbf{0.049} & \textbf{0.018} & \textbf{94.2} & 0.503 \\ 
$\gamma$ & -0.5 & I & -0.619 & -0.119 & \textbf{0.016} & 87.6 & \textbf{0.333}  \\ 
   & & M & -0.238 & 0.262 & 0.079 & 46.6 & 0.487 \\ 
   & & V & -0.480 & \textbf{0.020} & 0.024 & \textbf{95.4} & 0.618 \\ 
\hline
\end{tabular}
\caption{For sample size  $n=4000$, the true values and estimated mean, bias, mean squared error (MSE), coverage (Cov) and 95\% coverage interval width (CIW) for model parameters under different priors, with I using informative priors, V vague priors and M Misspecified priors.}
\label{tab: n4000}
\end{table}

\begin{table}[ht]
\centering
\begin{tabular}{p{1.3cm}cccrrrrr}
    \toprule
 & S. & True & Model & Mean & Bias & MSE & Cov. \% & CIW.  \\ \midrule
\multirow{8}{\linewidth}{2 year survival}  
&\multirow{4}{*}{16} &\multirow{4}{*}{0.788} & I & 0.778 & -0.010 & $<$0.001 & 71.8 & 0.029\\ 
&  & & M & 0.789 & 0.001 &$<$0.001 & \textbf{95.0} & \textbf{0.028}   \\ 
&  &  & V & 0.787 & \textbf{-0.001} & \textbf{$<$0.001} & 94.2 & 0.030  \\ 
&  & & PI & 0.797 & 0.008 & $<$0.001 & 79.8 & 0.029 \\ 
\cmidrule{2-2}
& \multirow{4}{*}{18} & \multirow{4}{*}{0.855} & I & 0.872 & 0.017 & $<$0.001& 55.4 & \textbf{0.036}  \\ 
&  & & M & 0.845 & -0.010 & \textbf{$<$0.001} & 86.4 & 0.044\\ 
&  & & V & 0.850 & \textbf{-0.005} & $<$0.001 & \textbf{92.2} & 0.046  \\ 
&  & & PI & 0.818 & -0.028 & 0.001 & 30.4 & 0.040\\ 
\cmidrule{1-2}
\multirow{8}{\linewidth}{5 year survival}  
&\multirow{4}{*}{16} &\multirow{4}{*}{0.755} & I & 0.741 & -0.015 & $<$0.001 & 52.4 & \textbf{0.030}  \\ 
&  & & M  & 0.747 & -0.008 & $<$0.001 & 80.4 & 0.032 \\ 
&  & & V & 0.757 & \textbf{0.001} & \textbf{$<$0.001} & \textbf{95.0} & 0.032  \\ 
&  & & PI & 0.762 & 0.007 & $<$0.001 & 83.2 & 0.030\\ 
\cmidrule{2-2}
& \multirow{4}{*}{18} &\multirow{4}{*}{0.763} & I & 0.810 & 0.047 & 0.002 & 0.8 & \textbf{0.039} \\ 
&  & & M & 0.773 & 0.010 & $<$0.001 & 90.2 & 0.055  \\ 
&  & & V & 0.767 & \textbf{0.004} & \textbf{$<$0.001} & \textbf{93.6} & 0.057  \\ 
&  & & PI & 0.833 & -0.070 & 0.005 & 0.0 & 0.043  \\ 
\cmidrule{1-2}
\multirow{8}{\linewidth}{10 year survival} 
&\multirow{4}{*}{16} &\multirow{4}{*}{0.755} & I  & 0.740 & -0.015 &$<$0.001 & 52.4 & \textbf{0.030}   \\ 
&  & & M & 0.746 & -0.009 & $<$0.001 & 78.0 & 0.032  \\ 
&  &  & V & 0.756 & \textbf{0.001} & \textbf{$<$0.001} & \textbf{95.2} & 0.032   \\ 
&  & & PI & 0.688 & -0.100 & 0.010 & 0.0 & 0.035 \\ 
\cmidrule{2-2}
& \multirow{4}{*}{18} & \multirow{4}{*}{0.756} & I & 0.804 & 0.048 & 0.002 & 0.2 & \textbf{0.040} \\ 
&  & & M & 0.767 & 0.012 &$<$0.001 & 87.2 & 0.056   \\ 
&  & & V & 0.761 & \textbf{0.005} & \textbf{$<$0.001} & \textbf{93.4} & 0.059  \\ 
&  & & PI & 0.730 & -0.026 & 0.001 & 52.4 & 0.056\\ 
\end{tabular}
\caption{For sample size $n=4000$, the true values and estimated mean, bias, mean squared error (MSE), coverage (Cov) and 95\% coverage interval width (CIW) for survival probability at time $t=2$, $t=5$ and $t=10$. Models compared are PI (Prevalence Incidence), I a Prevalence Incidence Cure model using informative priors, V a Prevalence Incidence Cure model using vague priors and M a Prevalence Incidence Cure model using Misspecified priors.}
\label{Tab: n4000Survproba}
\end{table}

\begin{table}
\centering
\begin{tabular}{cc|ccccccccccccc}
    \toprule
 & S. & True & Model & Mean & Bias & MSE & Cov. \% & CIW.  \\ \midrule
\multirow{8}{*}{Prev.} 
&\multirow{4}{*}{16} &\multirow{4}{*}{0.169} & I & 0.166 & \textbf{-0.003} & $<$0.001 & 94.0 & 0.026   \\ 
&  & & M & 0.164 & -0.005 & \textbf{$<$0.001} & 87.2 & \textbf{0.026} \\ 
&  & & V & 0.168 & $<$0.001 & $<$0.001 & \textbf{95.2} & 0.027 \\ 
&  & & PI & 0.190 & 0.021 & 0.001 & 18.0 & 0.030  \\ 
\cmidrule{2-2}
& \multirow{4}{*}{18} & \multirow{4}{*}{0.081} & I & 0.085 & 0.004 & \textbf{$<$0.001} & \textbf{92.6} & \textbf{0.034} \\ 
&  & & M & 0.094 & 0.013 & $<$0.001 & 70.2 & 0.036 \\ 
&  & & V & 0.085 & \textbf{0.004} & $<$0.001 & 92.2 & 0.036 \\ 
&  & & PI & 0.099 & 0.018 & $<$0.001 & 55.0 & 0.041  \\ 
\cmidrule{1-2}
\multirow{6}{*}{Cure} 
&\multirow{3}{*}{16} &\multirow{3}{*}{0.755} & I & 0.740 & -0.015 & $<$0.001 & 52.4 & \textbf{0.030}   \\ 
&  & & M & 0.746 & -0.009 & $<$0.001 & 78.0 & 0.032  \\ 
&  & & V & 0.756 & \textbf{0.001} & \textbf{$<$0.001} & \textbf{95.2} & 0.032 \\ 
\cmidrule{2-2}
& \multirow{3}{*}{18} & \multirow{3}{*}{0.756} & I & 0.804 & 0.048 & 0.002 & 0.2 & \textbf{0.040} \\ 
&  & & M & 0.767 & 0.011 & $<$0.001 & 87.2 & 0.056  \\ 
&  & & V & 0.761 & \textbf{0.005} & \textbf{$<$0.001} & \textbf{93.4} & 0.059\\ 
\hline
\end{tabular}
\caption{For sample size $n=4000$, the true values and estimated mean, bias, mean squared error (MSE), coverage (Cov) and 95\% coverage interval width (CIW) for cure and prevalence under strain 16 and strain 18. Models compared are PI (Prevalence Incidence), PIC (Prevalence Incidence Cure) with I using informative priors, V vague priors and M Misspecified priors.}
\label{Tab: n4000Survprobb}
\end{table}


\end{document}